\documentclass[12pt]{article}
\usepackage{amsmath}
\usepackage{amsfonts}
\usepackage{mathrsfs}
\usepackage{mathdots}
\usepackage{mathtools}
\usepackage[backend=bibtex,
style=authoryear,
sorting=nty
]{biblatex}
\usepackage[titletoc,title]{appendix}
\usepackage{chngcntr}
\usepackage{apptools}

\newtheorem{theorem}{Theorem}
\newtheorem{lemma}{Lemma}

\newcommand{\R}{\mathbb{R}}

\newcommand{\Vt}{\boldsymbol{v}}

\newcommand{\bX}{\boldsymbol{X}}

\newcommand{\bH}{\boldsymbol{H}}

\newcommand{\bZ}{\boldsymbol{Z}}
\newcommand{\bhX}{\hat{\boldsymbol{X}}}

\newcommand{\bep}{\boldsymbol{\epsilon}}

\newcommand{\bhep}{\hat{\boldsymbol{\epsilon}}}

\newcommand{\btheta}{\boldsymbol{\theta}}

\newcommand{\bPhi}{\boldsymbol{\Phi}}

\newcommand{\bXtheta}{\boldsymbol{X}_{\theta}}

\newcommand{\lag}{\textsc{lag}}
\newcommand{\T}{\mathscr{T}}
\newcommand{\mL}{\mathscr{L}}
\newcommand{\mLb}{\bar{\mathscr{L}}}
\newcommand{\bOmega}{\boldsymbol{\Omega}}
\newcommand{\bSigma}{\boldsymbol{\Sigma}}
\newcommand{\ThetaT}{\Theta_T}
\newcommand{\pdf}{\boldsymbol{pdf}}
\newcommand{\bsh}{\boldsymbol{h}}
\newcommand{\bsu}{\boldsymbol{u}}
\newcommand{\Kbar}{\bar{K}}
\DeclareMathOperator{\Tr}{Tr}

\DeclarePairedDelimiter\floor{\lfloor}{\rfloor}

\title{An Explicit Formula for Likelihood Function for Gaussian Vector Autoregressive Moving-Average Model Conditioned on Initial Observables with Application to Model Calibration.}
\author{Du Nguyen \\ Statistical Alpha Fund Management LLC \\du.nguyen@statisticalalpha.com}
\begin{document}
\maketitle
\abstract{We derive an explicit formula for likelihood function for Gaussian VARMA model conditioned on initial observables where the moving-average (MA) coefficients are scalar. For fixed MA coefficients the likelihood function is optimized  in the autoregressive variables $\Phi$'s by a closed form formula generalizing regression calculation of the VAR model with the introduction of an inner product defined by MA coefficients. We show the assumption of scalar MA coefficients is not restrictive and this formulation of the VARMA model shares many nice features of VAR and MA model. The gradient and Hessian could be computed analytically. The likelihood function is preserved under the root invertion maps of the MA coefficients. We discuss constraints on the gradient of the likelihood function with moving average unit roots. With the help of FFT the likelihood function could be computed in $O((kp+1)^2T +ckT\log(T))$ time. Numerical calibration is required for the scalar MA variables only. The approach can be generalized to include additional drifts as well as integrated components. We discuss a relationship with the Borodin-Okounkov formula and the case of infinite MA components.}
\section{Introduction}
The main result of this paper is the following:

\begin{theorem} The conditional log-likelihood function of a $k$-dimension vector autoregressive moving-average model (VARMA)
\begin{equation}
X_t = \mu +  X_{t-1} \Phi_1+ X_{t-2} \Phi_2+\cdots+ X_{t-p} \Phi_p +\epsilon_t + \theta_1 \epsilon_{t-1}+\cdots+\theta_q \epsilon_{t-q}
\end{equation}
conditioned on the first $p$ observations ($X_1,\cdots, X_p$) of the $T+p$ observations $X_1,\cdots,X_p,X_{p+1},\cdots X_{T+p}$ with $\theta_1,\cdots \theta_q$ are scalars
is given by the formula
\begin{multline}
\mathscr{L}(\btheta, \mu, \bPhi,\bOmega,X_{p+1}\cdots X_{T+p}|X_1\cdots X_p) = -\frac{Tk}{2}\log(2\pi) -\frac{T}{2}\log(\det(\bOmega)) \\- {k/2} \log(\det(\lambda^{\prime}\lambda+I_q)) 
-\frac{1}{2}\Tr(\bZ^{\prime}\Theta_T^{-1\prime}K(\btheta,T)\Theta_T^{-1} \bZ\bOmega^{-1}))
\label{eq:llk1}
\end{multline}
where $\theta_0=1$, 
$\btheta =(\theta_1,\cdots,\theta_q)$,
$\Phi=(\Phi_1,\cdots,\Phi_p)$,
$\bOmega$ is the covariance matrix of the $i.i.d.$ Gaussian random variables $\epsilon_i$'s. Here:
\[ \bZ =  \bX-\mu - L \bX \Phi_1-...- L^p \bX \Phi_p\]

\[\bX= \begin{pmatrix} X_{p+1}\\ \cdots \\ X_{T+p}\end{pmatrix}\]
of size $T\times k$.

\[L^i\bX = \begin{pmatrix}X_{p-i+1}\\ \cdots \\ X_{T+p-i}\end{pmatrix}\]

\begin{equation}
\Theta_T = 
\begin{pmatrix}
 \theta_0 & 0 & \cdots & 0 & 0 & 0\\
\theta_1 & \theta_0 & 0 & \cdots &  0 & 0 \\
\vdots& \vdots& \vdots & \vdots & \vdots & \vdots\\
\theta_{q-1} & \theta_{q-2} & \cdots & \cdots & 0 & 0\\
\theta_q & \theta_{q-1} & \theta_{q-2} & \cdots & 0 & 0\\ 
0 & \theta_q & \theta_{q-1} &\cdots & 0 & 0\\
\vdots & \vdots & \vdots & \vdots & \vdots & \vdots \\
0 & 0& 0 & \cdots & 0 & \theta_0
\end{pmatrix}
\end{equation}
is of size $T\times T$.
\begin{equation}
\lambda = \ThetaT^{-1}\Theta_{*;T-q}
\end{equation}
is of size $T\times q$.
where
\begin{equation}
\Theta_* = 
\begin{pmatrix}
\theta_{q}& \theta_{q-1}&\cdots &\cdots & \cdots & \theta_1\\
0 &  \theta_{q}& \theta_{q-1} &\cdots &\cdots& \theta_2\\
0 & 0 & \theta_{q} & \theta_{q-1} & \cdots& \theta_3 \\
\vdots &\vdots & \vdots &\vdots & \vdots\\
0 & 0  & \cdots  &\cdots & 0 &\theta_q\\
\end{pmatrix}
\end{equation}
is of size $q\times q$ and
\[
\Theta_{*;T-q} = \begin{pmatrix}\Theta_* \\ 0_{T-q,k} \end{pmatrix} 
\]

\begin{equation}
K = K_T K(\btheta,T) = I_T -\lambda [\lambda^{\prime} \lambda+I_q]^{-1}\lambda^{\prime}  = (I_T+\lambda\lambda^{\prime})^{-1} 
\end{equation}
The optimal value is obtained at
\begin{equation}
\begin{pmatrix}\mu \\ \Phi_1\\ \Phi_2\\ \vdots \\ \Phi_p \end{pmatrix}_{opt} =  (\bX_{\theta,\lag}^{\prime}K\bX_{\theta,\lag})^{-1} \bX_{\theta,\lag}^{\prime}K\bXtheta
\end{equation}
where:
\begin{equation}
\bXtheta = \ThetaT^{-1}\bX
\end{equation}
\[ \bX_{\theta,\lag}= \begin{pmatrix}\ThetaT^{-1}1 & \ThetaT^{-1}\bX &\ThetaT^{-1}L\bX &\cdots &\ThetaT^{-1}L^p\bX
\end{pmatrix}
\]
and
\begin{equation}
\bOmega_{opt}(\theta) = \frac{1}{T}
[\bXtheta^{\prime} K\bXtheta - 
 \bXtheta^{\prime} K \bX_{\theta,\lag} (\bX_{\theta,\lag}^{\prime}K\bX_{\theta,\lag})^{-1} \bX_{\theta,\lag}^{\prime}K \bXtheta]
\end{equation}
$\bOmega_{opt}$ is positive semi-definite regardless of sample values of $X$ and choice of $\btheta$. With these values of $\Phi_{opt}$ and $\bOmega_{opt}$, (\ref{eq:llk1}) is reduced to
\begin{multline}
\mLb(\btheta,X_{p+1}\cdots X_{T+p} | X_1\cdots X_p) = -\frac{Tk}{2}\log(2\pi)-\frac{T}{2}\log(\det(\bOmega_{opt}(\theta)))-\\
\frac{k}{2}\log(\det(\lambda^{\prime}\lambda+I_q))-\frac{Tk}{2}
\label{eq:llk2}
\end{multline}
Futher, set 
\begin{equation}
\Sigma_T =\begin{pmatrix} \gamma_0 & \gamma_1 & \gamma_2 &\cdots &\gamma_q& 0 & \cdots & 0 \\
\gamma_1 & \gamma_0 & \gamma_1 & \gamma_2 & \cdots & \gamma_q &\cdots & 0 \\
\vdots&\ddots &\ddots &\ddots &\ddots &\ddots &\ddots &\vdots \\
\vdots&\ddots &\ddots &\ddots &\ddots &\ddots &\ddots &\vdots \\
\vdots&\ddots &\ddots &\ddots &\ddots &\ddots &\ddots &\vdots \\
\vdots&\ddots &\ddots &\ddots &\ddots &\ddots &\ddots &\vdots \\

0 &\cdots & 0 &\gamma_q &\cdots & \gamma_1 & \gamma_0 &\gamma_1 \\
0 &\cdots & 0 & 0 &\gamma_q &\cdots & \gamma_1 & \gamma_0
\end{pmatrix}
\end{equation}
with
\begin{equation} \gamma_l = \left\{ \begin{array}{l l} (\theta_l +\theta_1\theta_{l+1}+\theta_2\theta_{l+2}\cdots + \theta_{q-l}\theta_q) & \text{for }l=0,1,\cdots,q \\
0 & \text{for }l > q
\end{array} \right .
\label{eq:gamma}
\end{equation}

Then
\begin{equation}
\Sigma_T^{-1} = \Theta^{-1\prime}_T K(\btheta,T)\Theta^{-1}_T
\label{eq:sigmatheta1}
\end{equation}
or 
\begin{equation}
\Sigma_T = \Theta_T K(\btheta, T)^{-1}\Theta_T^{\prime}
\label{eq:sigmatheta2}
\end{equation}
also we have
\begin{equation}
\det(\lambda^{\prime}\lambda +I_q) = \det(\Sigma_T)=\frac{1}{\det(K(\btheta, T))}
\label{eq:sigmatheta3}
\end{equation}
\end{theorem}

$\Sigma_T$ is the well-known concentrated covariance matrix in the study of MA$(q)$ process associated with $\btheta$ (normalized to standard deviation of noise equals 1). We note this likelihood function is conditional only on the $p$ observations of $X$, and not on the initial error estimates $\epsilon$ in contrast with the typical conditional sum of squares (CSS) approach. In particular, for VMA models with scalar $\btheta$, the formula gives an exact likelihood formula. For scalar MA models, the formula for the likihood function in term of $\bSigma_T$ is the same as those found in standard text books, e.g. \parencite{BJ} or \parencite{Hamilton}. The determinant of $\Sigma_T$ in (\ref{eq:sigmatheta3}) is one studied in the strong Szeg\"o limit theorem and the Borodin-Okounkov's determinant formula \parencite{GeCase,BO,BasorWidom} in the theory of Toeplitz operators. While we use the Szeg\"o limit theorem to express the large $T$ limit of the determinant in close form, we do not need to use the Fredholm determinant result in this paper but just mention the context that the determinants in (\ref{eq:sigmatheta3}) have appeared elsewhere in the literature. The construction of $\lambda$ and $\Kbar$ seems new but we could not be sure it has not appeared in the multiple proofs of the Borodin-Okounkov formula. $K$ is related to the matrix $A$ in the second proof of the Borodin-Okounkov's formula in \parencite{BasorWidom}. 

This decomposition permits more effective calculations of the likelihood function when $T$ is large. We note (\ref{eq:sigmatheta1}), (\ref{eq:sigmatheta2}), (\ref{eq:sigmatheta3}) are purely algebraic, depending only on $\theta$ and $T$. To verify them by hand for a few $\theta$ and $T$ would be interesting exercises. For example, with $q=1$ (\ref{eq:sigmatheta3}) shows the determinant of $\Sigma_T$ is $1+\theta_1^2+\cdots\theta_1^{2T}$, a result well-known in most time series text books.

Using $\Sigma_T$, we can rewrite :
\begin{equation}
\begin{pmatrix}\mu \\ \Phi_1\\ \Phi_2\\ \vdots \\ \Phi_p \end{pmatrix}_{opt} =  (\bX_{\lag} \Sigma_T^{-1}\bX_{\lag})^{-1} \bX_{\lag}^{\prime}\Sigma_T^{-1}\bX
\label{eq:ylk}
\end{equation}
with 
\[ \bX_{\lag}= \begin{pmatrix}1 & \bX & L\bX &\cdots &L^p\bX
\end{pmatrix}
\]

\begin{equation}
\Omega_{opt}(\theta) = \frac{1}{T}
[\bX^{\prime} \Sigma_T^{-1}\bX -
 \bX^{\prime} \Sigma_T^{-1} \bX_{\lag} (\bX_{\lag}^{\prime}\Sigma_T^{-1}\bX_{\lag})^{-1} \bX_{\lag}^{\prime}\Sigma_T^{-1} \bX]
\end{equation}

We will use the notations:
\[\btheta(L) = 1 + \theta_1 L + \cdots + \theta_q L^q\]
\[\bPhi(L) = 1 - \Phi_1 L - \cdots - \Phi^p L^q\]

The condition that $\btheta$ is scalar is not restrictive, in the sense that given a system with matrix $\Theta(L)$, we can transform it into one with the same transfer function and scalar MA components. The reverse case, scalar $\Phi$ is already well-known (for example in Gilbert realization in Linear System literature - for a time series treatment see \parencite{Aoki}) - chapter 4.

Let us recall that argument. If $N(L)$ and $D(L)$ are two square matrix polynomials. We can write $T(L) = N(L)^{-1}D(L)$ as a matrix with rational functions entries $t_{ij}$. Assume all entries $t_{ij}(L) = nt_{ij}(L)/dt_{ij}(L)$ with $nt_{ij}$ and $dt_{ij}$ are relative prime. Take the least common multiple (lcm) polynomial of all the denominators polynomial entries $dt_{ij}$, call it $d(L)$. $d(L)T(L)=\Phi(L)$ is a polynomial matrix, so we have proved $T$ could be written as $\frac{1}{d(L)}\Phi(L)$ with $d$ scalar and $\Phi$ polynomial. Alternatively, and this is what we will use in our simulation result, is to write 
\[T(L) = N(L)^{-1}N_A(L)^{-1}N_A(L)D(L) =(\det(N(L))^{-1}N_A(L)D(L)\]
where $N_A(L)$ is the adjugate matrix of $N$. The last expression is of the desired form with $\btheta = \deg(N(L))$. We note if $\text{deg}(N) \geq \text{deg}(D)$ and $D(L), N(L)$ comes from a minimal realization (via the Kronecker index approach for example) then $\text{deg}(\det(N))$ is the McMillan degree $\delta(T)$ of the process. We will come back to this discussion in the later section on calibration.

The likelihood formula is valid for any sample size, with no restriction on location of roots of $\btheta$. However for invertible $\btheta(L)$, the terms of $\Theta_T^{-1}$ converges as $T$ increase. Here, we apply an observation of \parencite{HansenSargent} that we can adjust the transfer function by Blaschke product terms but still preserve the autocovariance function (this is just the trick to replace a root of $\btheta$ by its inverse.) Further calculations show \parencite{Hamilton} that inverting of a root results in multiplying $\Sigma_T$ with the square of that root. In section \ref{appxIR} we will examine how different components in the above theorem transform under root inverting and verify that the likelihood function above is invariant under the operation of inverting any number of roots. Therefore we can restrict ourselves to working with models with invertible $\btheta$ only.

There are several advantages in using the above likelihood formula for model calibration. First of all, only $q$ variables need to be optimized numerically, the $\theta$ variables. Secondly, we only need to "throw away" only the first $p$ observations. This is in contrast with the CSS method for typical MA models. If the optimization path get to a root of $\theta(L)$ close to $1$, coefficients take a long time to decay so a typical CSS needs to throw away many terms before the forecast become stable. Finally, it also compare favorably with the Kalman filter calibration approach. In the multivariate case calibration usually requires Kronecker indices to reduce rank, otherwise the number of variables involved would be $pk^2 +q$. For a process with Kronecker indices $(d_1,\cdots,d_i,\cdots,d_k)$ with McMillan degree $m=\sum d_i$ the number of parameters in a typical estimate is \parencite{Tsay}
\[ m(k+1) +  \sum_{j=1}^k [\sum_{i < j} \min\{d_j+1,d_i\} + \sum_{j>i}\min \{d_j,d_i\}] \]
With our approach, we only have $m$ variables to be optimized numerically while the rest are computed via regression. Also gradients are harder to compute in the traditional approach. Finally we only need to estimate before hand the McMillan degree as an upper bound for $q$, not the whole set of Kronecker indices. However we advocate further test to reduce the number of non-zero coefficients to simplify the model.

Our approach is hybrid. We use exact likelihood to for MA terms and try to take advantage of the regression formula for AR terms. In effect, it allows for an efficient search for the scalar polynomial $\btheta(L)$ that removes the moving average complexity and leave us with AR data where we can apply regression. We can think of this approach as smoothing then regressing, where we have an efficient algorithm to search for smoothing parameters.

We will see in subsequent sections that this conditional likelihood could be computed relatively fast, with the use of convolution algorithm together with Fast Fourier Transform, allowing us to attack very large sample size. Secondly, exact gradient of the likelihood function is computed with ease, resulting in an efficient optimization algorithm. The C++ and R codes developed by the author implement this algorithm. Conceptually, even the evaluation of the Hessian could be done in reasonable time. However for immediate applications the gradient seems sufficient.
We note a few algorithms in exact likelihood estimation try to decompose $\Sigma_T$ to $LL^{\prime}$ form. In our approach instead of the standard Cholesky decomposition for $\Sigma_T$, we use the fact that $\Sigma_T^{-1}$ could be decomposed to sum of product of matrices that are triangular and Toeplitz ($\Theta_T^{-1}$) or have small number of rows or columns ($\lambda$ and $\bar{K}$.) We only need to do Cholesky decomposition of $\Kbar$ (of size $q\times q$) instead of a matrix of size $T\times T$.

Look at it another way, the theorem says that if $\btheta(L)$ is scalar, there exists an inner product defined by a {\it kernel} given by the positive-definite matrix $\Sigma_T^{-1}$. This inner product accounts for the moving average terms. Under this inner product the autoregresive term and likelihood function have simple format similar to the vector autoregressive (VAR) case. The inner product could be evaluated efficiently with the help of FFT and numerical calibration would need to be done on the $\btheta$ parameters only.

It is well-known that finite state space linear time invariant systems are exactly those having rational matrix transfer functions. So our result here could be understood as an explicit form of likelihood function for finite state Kalman filter in a MIMO system, conditioned on the first $p$ observations. We do see a possibility that this approach could be useful in calibrating Kalman filters in general.

\section{Proof of the theorem}
We start out with a general lemma
\begin{lemma}
Let $A$ be an arbitrary $T\times q$ matrix. Then
\begin{equation}
(I_T+AA^{\prime})^{-1} = I_T-A(I_q+A^{\prime}A )^{-1}A^{\prime}
\end{equation}
In particular the matrix on the right hand side has all eigenvalues in the closed unit disc.
\begin{equation}
\det(I_q + A^{\prime}A) = \det(I_T-A(I_q+A^{\prime}A )^{-1}A^{\prime})^{-1}
\end{equation}
\label{lem:wood}
\end{lemma}
This is a special case of Woodbury matrix identity:
\[   \left(A+UCV \right)^{-1} = A^{-1} - A^{-1}U \left(C^{-1}+VA^{-1}U \right)^{-1} VA^{-1} \]
and the Sylvester determinant's entity 
\[\det(I_q+AB) = \det(I_T +BA)\]
We note Woodbury matrix identity already has applications in Kalman filter update so we find it interesting but not quite surprising that it plays a core role in our formulation.

Let $Z_t$ be the time series defined by:
\begin{equation} Z_t = X_t- \mu- X_{t-1}\Phi_1  -\cdots- X_{t-p}\Phi_p = \epsilon_t + \theta_1 \epsilon_{t-1}+\cdots+\theta_q \epsilon_{t-q}
\label{eq:zeq}
\end{equation}

Assuming we have $n=T+p$ samples $X_{1},\cdots,X_p,X_{p+1},\cdots,X_{T+p}$ considered as rows of a matrix 

\[\bhX= \begin{pmatrix} X_1\\ \cdots \\ X_{T+p}\end{pmatrix}\]
of size $(T+p)\times k$.

Let 
\[
\bZ = \begin{pmatrix}Z_{p+1}\\ \cdots \\ Z_{T+p}\end{pmatrix}
\]

\[ \bep=\begin{pmatrix}\epsilon_{p+1}\\ \vdots \\ \epsilon_{T+p}\end{pmatrix} \]
\[ \bep_* = \begin{pmatrix} \epsilon_{p-q+1}\\\cdots\\\epsilon_p\end{pmatrix} \]

Then the equation (\ref{eq:zeq}) gives:
\begin{equation}
\bZ = \Theta_T \bep + \Theta_{*,T-q}  \bep_*
\end{equation}

We note $\Theta^{-1}_T$ could be constructed from the power series expansion of $\btheta(L)^{-1} = (\theta_0 +\theta_1 L+\cdots+\theta_pL^p)^{-1}$ via the Toeplitz map. Recall that for any integer $T>0$, the map $\T$ mapping a polynomial $(\theta_0 +\theta_1 L+\cdots+\theta_pL^p)$ to the matrix $\Theta_T$ above preserves addition, unit ($1$ is mapped to $I_T$), scalar multiplication and map polynomial multiplication to matrix multiplication. (In algebra language, it is a homomorphism from the matrix algebra of polynomial matrices $\R[L]$ to the algebra $M_T(\R)$ of$T\times T$ matrices). Because of this property, $\Theta_T^{-1}$ is the image of the truncated power series of $\btheta(L)^{-1} = (\theta_0 +\theta_1 L+\cdots+\theta_pL^p)^{-1}$ truncated at $T$ terms.

We solve for $\bep$ in term of $\bZ$ as:
\begin{equation}
\bep  = \Theta_T^{-1} \bZ - \Theta_T^{-1}\Theta_{*;T-q} \bep_*
\label{eq:bep}
\end{equation}
Set
\begin{equation}
\lambda = \Theta_T^{-1}\Theta_{*;T-q}
\end{equation}

We note that the $i$th-column of $\lambda$ could be constructed by truncating the first $T$ terms of the power series expansion $(\theta_{q-i}+ \theta_{q-i+1}L \cdots+\theta_{q}L^k)\boldsymbol{\theta}(L)^{-1}$.

Consider the vectorization that sends a $T\times k$ matrix to a $T\times k$ vector, where we expand the rows first:
\begin{equation}
v(A) = vec(A^{\prime})
\end{equation}

Let $\bhep$ be the $T+q$ matrix formed by adding the vector $\bep_*$ at the beginning of $\bep$: 
\[
\bhep=\begin{pmatrix}\epsilon_{p-q+1}\\ \cdots \\ \epsilon_1 \\ \cdots \\ \epsilon_{T+p}  \end{pmatrix}\]

From the  relation
\[ (B^T\otimes A) vec(X) = vec(AXB)\]
We have 
\[ v(\Theta\bep) = (\Theta\otimes I_k) v(\bep) \]

The covariance matrix for $v(\bhep)$ is a $(T+q) k\times (T +q)k$ matrix, with diagonal blocks of size $k\times k$ equal to $\bOmega$, and zero elsewhere. We will denote it by $\hat{\bOmega}_{T+q} = I_{T+q}\otimes \bOmega$. The join $\pdf$ of $\epsilon_{p-q+1},...,\epsilon_{T+p}$ is given by:
\begin{equation}
(2\pi)^{-(T+q)k/2}\det(\hat{\bOmega}_{T+q})^{-1/2}\exp(-\frac{1}{2}v(\bhep)^{\prime}\hat{\bOmega}^{-1}_{T+q}v(\bhep))
\end{equation}
which could be simplified to
\[
((2\pi)^k\det(\bOmega))^{-(T+q)/2}\exp(-\frac{1}{2}[v(\bep)^{\prime}\hat{\bOmega}^{-1}_T v(\bep) +
v(\bep_*)^{\prime}\hat{\bOmega}^{-1}_q v(\bep_*)]
)
\]
where $\hat{\bOmega}^{-1}_T$ and $\hat{\bOmega}^{-1}_q$ are diagonal block matrices $I_T\otimes \bOmega^{-1}$ and $I_q\otimes \bOmega^{-1}$ respectively.

Now we look for the marginal $\pdf$ with respect to $\bZ$, assuming we $\bep$ is related to $\bZ$ and $\bep_*$ by equation (\ref{eq:bep}). The approach of taking expectation with respect to initial terms is well-known, where $\bep_*$ are the initial terms. The $\pdf$ of $\bZ$ is
\[
((2\pi)^k\det(\bOmega))^{-(T+q)/2}\int_{\bep_*\in (\R^k)^q} e^{-\frac{1}{2}[v(\bep)^{\prime}\hat{\bOmega}^{-1}_T v(\bep) +
v(\bep_*)^{\prime}\hat{\bOmega}^{-1}_q v(\bep_*)]}\boldsymbol{d}\bep_*
\]
where 
\[ \boldsymbol{d}\bep_* = d\epsilon_{1,1}d\epsilon_{1,2}\cdots d\epsilon_{1,k}\cdots d\epsilon_{p,1}\cdots d\epsilon_{p,k}
\]
is the volume component of all the coordinates of $\bep_*$.

Expanding using (\ref{eq:bep}), the exponent could be written in the form: 
\[
\exp(-\frac{1}{2}[v(\bep_*)^{\prime}Av(\bep_*) + 2v(\bZ)^{\prime}Bv(\bep_*) +v(\bZ)^{\prime}Cv(\bZ) ]
\]

with 
\begin{equation}
A=(\lambda^{\prime} \otimes I_k) (I_T \otimes \bOmega^{-1}) (\lambda \otimes I_k) + I_q\otimes\bOmega^{-1} = (\lambda^{\prime}\lambda +I_q)\otimes \Omega^{-1}
\end{equation}

\begin{equation}
B= -(\Theta_T^{-1\prime}\otimes I_k )  (I_T\otimes\bOmega^{-1})(\lambda \otimes I_k) =-\Theta_T^{-1\prime}\lambda\otimes\bOmega^{-1}
\end{equation}

\begin{equation}
C= \Theta_T^{-1\prime} \Theta_T^{-1} \otimes \Omega^{-1}
\end{equation}

Here $A$ is a $qk\times qk$ matrix, $B$ is a $Tk\times qk $ matrix and $C$ is a $Tk\times Tk$ matrix. Using the formula
\[
\begin{split}
\int_{\bsu\in R^N} \exp(-\frac{1}{2}[\bsu^{\prime}A\bsu^* + 2\bsh^{\prime}B u^* + \bsh^{\prime}C \bsh ]) \boldsymbol{d}\bsu \\
= (2\pi)^{N/2} (\det(A))^{-1/2}\exp(-\frac{1}{2}[\bsh^{\prime}(C-BA^{-1}B)\bsh])
\end{split}
\]
with $N=kq$ is the dimension of $\bsu=\bep_*$ and
\[\det(A) = \det(\lambda^{\prime}\lambda+I_q)^k \det(\bOmega)^{-q} \]
we deduce:
\begin{equation}
\pdf(\bZ) = \frac{\exp(-\frac{1}{2}[v(\bZ)^{\prime}[C-BA^{-1}B^{\prime} ]v(\bZ)])}{((2\pi)^k\det(\bOmega))^{T/2}\det(\lambda^{\prime}\lambda+I_q)^{k/2}} 
\label{eq:pdfZ1}
\end{equation}

The exponent is quadratic in $\bZ$. Note
\[  C-BA^{-1}B^{\prime} = \Theta_T^{-1\prime}[I_T -\lambda [\lambda^{\prime} \lambda+I_q]^{-1}\lambda^{\prime}]\Theta_T^{-1} \otimes \Omega^{-1}
\]
By lemma \ref{lem:wood} $I_T -\lambda [\lambda^{\prime} \lambda+I_q]^{-1}\lambda^{\prime}$ is positive definite. The following lemma is well-known in vectorization: 
\begin{lemma}
For any two matrices of the same size $M$ and $N$, 
\begin{equation} v(M)^{\prime}v(N) = \Tr(M^{\prime}N) 
\end{equation}
In particular, if $H$ and $\bOmega$ are symmetric of size $T\times T$ and $k\times k$ respectively then if $X$ is a matrix of size $T\times k$ we have
\[ v(X)^{\prime} (H \otimes \bOmega) v(X) =v(HX)^{\prime}v(X\bOmega)= \Tr( X^{\prime} H X \bOmega) \]
\label{lem:traceKro}
\end{lemma}
This provides a connection between $vec$ and $\Tr$.
Set
\[
\bar{K} = \bar{K}(\btheta) = \lambda^{\prime} \lambda+I_q
\]
Then $\bar{K}$ is an $q\times q$ matrix.
\[
K = K(\btheta,T) = I_T -\lambda \bar{K}^{-1}\lambda^{\prime} = I_T -\lambda [\lambda^{\prime} \lambda+I_q ]\lambda^{\prime}
\]

$K$ is a $T \times T$ matrix. Then

\begin{equation}
pdf(\bZ |X,\btheta,\bPhi,\bOmega) = \frac{\exp(-\frac{1}{2}\Tr(\bZ^{\prime}\Theta_T^{-1\prime}K(\btheta,T)\Theta_T^{-1} \bZ\bOmega^{-1}))}{((2\pi)^k\det(\bOmega))^{T/2}\det(\lambda^{\prime}\lambda+I_q)^{k/2}}  
\label{eq:pdfZ}
\end{equation}
From here we have proved (\ref{eq:llk1}).
Let us now consider the partial optimization problem in $\bPhi$ and $\bOmega$ given $\btheta$. Substitute
\[ \bZ =  \bX-\mu - L \bX \Phi_1-...- L^p \bX \Phi_p\]
in equation (\ref{eq:pdfZ}), the problem is to find $\Phi_i$ minimizing:
\begin{multline}
\Tr((\bX-\mu -  L \bX \Phi_1 -...- L^p \bX\Phi_p)^{\prime}\Theta^{-1\prime} K(\btheta,T)\Theta_T^{-1}\\
(\bX-\mu -L \bX \Phi_1  -...- L^p \bX \Phi_p)\bOmega^{-1})=\\
\Tr((\Theta^{-1}\bX-\Theta_T^{-1}\mu -  \Theta^{-1}L \bX \Phi_1 -...- \Theta^{-1}L^p \bX\Phi_p)^{\prime} K(\btheta,T)\\
(\Theta^{-1}\bX-\Theta^{-1}\mu -\Theta^{-1}L \bX \Phi_1  -...-\Theta^{-1} L^p \bX \Phi_p)\bOmega^{-1})
\label{eq:quadtr}
\end{multline}

Since the expression is quadratic in $\mu$ and $\Phi_i$'s they could be optimized via linear regression with a modified inner product. We form the matrix $\bX_{\theta,\lag}$ of size $T\times (k\times (1+p))$ as
\[(\Theta_T^{-1}1| \Theta_T^{-1}\bX|\Theta_T^{-1}L\bX | \cdots | \Theta_T^{-1}L^p\bX ).\] 
If we do not include the constant term $\mu$ we could exclude the block $\Theta_T^{-1} 1$. Set
\begin{equation}
\bXtheta = \Theta_T^{-1}\bX
\end{equation}

Then
\[ \begin{pmatrix}\mu \\ \Phi_1\\ \Phi_2\\ \vdots \\ \Phi_p \end{pmatrix}_{opt} =  (\bX_{\theta,\lag}^{\prime}K\bX_{\theta,\lag})^{-1} \bX_{\theta,\lag}^{\prime}K\bXtheta
\]
is the optimum choice. This is proved in lemma \ref{lem:cauchy}. We note that it does not depend on $\Omega$.

With this choice of $\Phi_i$ the minimal value of the quadratic form (\ref{eq:quadtr}) above is
\begin{multline}
\Tr(( \bXtheta)^{\prime} K\bXtheta\Omega^{-1} - \\
\Tr(( \bXtheta)^{\prime} K \bX_{\theta,\lag} (\bX_{\theta,\lag}^{\prime}K\bX_{\theta,\lag})^{-1} \bX_{\theta,\lag}^{\prime}K  \bXtheta\Omega^{-1})
\end{multline}
Similar to the maximum likelihood argument for the VAR model, an argument using Jacobian formula relating derivative of $\det$ and $\Tr$ shows the choice of $\Omega$ that minimize the log likelihood is:
\[
\bOmega_{opt}(\theta) = \frac{1}{T}
(\bXtheta)^{\prime} K\bXtheta - \\
( \bXtheta)^{\prime} K \bX_{\theta,\lag} (\bX_{\theta,\lag}^{\prime}K\bX_{\theta,\lag})^{-1} \bX_{\theta,\lag}^{\prime}K \bXtheta
\]
Finally, with that value of $\bOmega$, the matrix inside the trace expression is simply $T.I_k$ and hence the trace is $Tk$. The conditional log-likelihood is:
\[
\begin{split}
\mLb(\btheta) = -\frac{Tk}{2}\log(2\pi)-\frac{T}{2}\log(\det(\bOmega_{opt}(\theta))) \\-\frac{k}{2}\log(\det(\lambda^{\prime}\lambda+I_q))-\frac{Tk}{2}
\end{split}
\]
We note the formulas appearing here look very much like regular regression/ covariance formulas, but with the inner product is given by $\Theta_T^{-1\prime}K(\theta,T)\Theta_T^{-1}$.

Let us discuss the relation connecting $\Sigma_T$ and $\Theta_T^{-1}$ and $K(\theta,T)$. This is purely an algebraic equality involving only $\btheta$. We observe that the likelihood function, in case of a pure moving average with  scalar $\theta$ is reduced to the scalar MA(q) model, tensoring with $\bOmega$. Comparing (\ref{eq:llk1}) in case $k=1,p = 0,\bOmega = \sigma^2, \mu=0 $
\[
\begin{split}
\mL(\btheta, X_{1}\cdots X_{T}) = -\frac{T}{2}\log(2\pi) -\frac{T}{2}\log(\det(\bOmega)) \\- \frac{1}{2} \log(\det(\lambda^{\prime}\lambda+I_q)) 
-\frac{1}{2}(\bX^{\prime}\Theta_T^{-1\prime}K(\btheta,T)\Theta_T^{-1} \bX\bOmega^{-1})
\end{split}
\]
with the known formula for MA(q) for example (5.5.5) in \parencite{Hamilton} (note $\Sigma_T$ in our notation is $\sigma^{-2}\bOmega$ in that reference's ): 
\[ \mL(\btheta) = -\frac{T}{2}\log(2\pi)-\frac{1}{2}\log(\det(\Sigma_T)) -\frac{1}{2}(\bX^{\prime}\Sigma_T^{-1}\bX) \]
we get the required equation.
\section{Szeg\"o's limit for MA($q$)}
For MA($1$) it is well known that the large $T$ limit of $\det(\Sigma_T^{-1}) =\det(K_T)$ is just $1-\theta_1^2$. As this determinant appears in the likelihood function, it would be natural to ask if a similar result hold in general. It turns out that that the large $T$ limit of the determinant is always a polynomial.

The strong Szeg\"o's limit theorem (\parencite{Szego, Bingham2012,BasorWidom}) shows how to compute the large $T$ limit for determinants for truncated Toeplitz matrices arising from certain analytic functions on the unit disc.
For Toeplitz matrix generated from rational functions (essentially general VARMA case) the limit is known under state space representation for example in \parencite{GOHBERG198724}. \parencite{KrRosen} also mentioned the theorem.
For the case MA($q$) the expression is very simple but we could not locate a reference so let us state:
\begin{theorem}
If $\btheta(L) = \prod_{i=0}^q (1-\lambda_iL)$ is invertible then
\begin{equation}
\lim_{T\to \infty} \det(\Sigma_T)^{-1} = (\sum_{i=0}^q \theta_i) ( \sum_{i=0}^q (-1)^i\theta_i) \prod_{1\leq i<j\leq q}(1-\lambda_j \lambda_j)^2
\end{equation}
The last term is a symmetric polynomial in $\lambda_i$'s so it could also be expressed as a polynomial in $\theta_i$'s
\end{theorem}
Apply Szeg\"o's limit theorem for the function $a(L) = \btheta(L)\btheta(L^{-1})$ we have 
\[
\lim_{T\to \infty} \det(\Sigma_T) = \exp(\sum_{k=1}^{\infty} k (\log(a))_k^2)
\]
$(\log(a))_k$ means we take the $kth$ coefficients of the Laurent expansion of $\log(a)$. But we have
\[\log(a(L)) = \sum_{i=1}^q\log(1-\lambda_iL)  +\sum_{i=1}^q \log(1-\lambda_iL^{-1})
\]
so we see easily:
\[
\log(\btheta(L))_k = \sum_{i=1}^q \frac{ \lambda_i^k}{k}
\]
So the exponent term is
\[
\sum_{k=1}^{\infty}  \frac{1}{k} (\sum_{i=1}^q \lambda_i^k)^2 = \sum_{i=1}^q \sum_k \frac{ \lambda_i^{2k}}{k} + 2\sum_{0\leq i<j \leq q}\sum_k\frac{(\lambda_i\lambda_j)^k}{k}
\]
and the limit of $\Sigma_T$ is
\[ \prod(1-\lambda_i^2)^{-1}\prod(1-\lambda_i \lambda_j)^{-2} = \prod(1-\lambda_i)^{-1}\prod(1+\lambda_i)^{-1} \prod(1-\lambda_i\lambda_j)^{-2} \]
which is what we have to prove. We can compute easily the expression for $\prod(1-\lambda_i\lambda_j)$ for small $q$.  The following table summarize $\lim_{T\to \infty} \Sigma_T^{-1}$ up to $q=3$.
\begin{center}
\begin{tabular}{ |c|c| } 
 \hline
 $q=1$ & $ 1-\theta_1^2$\\ 
$q=2$ & $(1-\theta_1^2)(1-\theta_2)$ \\ 
 $q=3$ &  $(1-\theta_1^2)(1-\theta_2+\theta_1\theta_3-\theta_3^2)$ \\ 
 \hline
\end{tabular}
\end{center}

For higher $q$, the polynomial expressed in term of $\theta$'s expands to a large number of monomial terms so it is simpler to evaluate in term of $\lambda_i$'s. We will see below they are the same polynomials that enforce invertibility condition for small $q$. Note that $\bar{K}_T$ is a $q\times q$ matrix so we can also attempt to compute its limit directly. While we could prove in large T limit, entries of $\bar{K}_T$ are rational functions in the $\theta_i$'s, the explicit expressions for the entries are complicated so the fact that the determinant is the reciprocal of a polynomial is interesting. 

A similar calculation could also be done for the covariance matrix in the invertible-stable ARMA($p,q$) case, the result involves roots of both the numerator and denominator of $\phi(L)/\theta(L)$ (we assume $\phi(0)= \theta(0)=1$) denoting them by $\mu_i^{-1}$ and $\lambda_j^{-1}$ with the assumption $|\mu_i| <1, |\lambda_j| < 1\forall i,j$.  The limit formula is:

\begin{multline}
\lim_{T\to \infty} \det(\Sigma_T) = 
\prod(1-\mu_i^2)^{-1}\prod(1-\mu_i \mu_{l})^{-2} \\
\prod(1-\lambda_j^2)^{-1}\prod(1-\lambda_j \lambda_m)^{-2} 
\prod(1-\mu_i\lambda_j )^{2} 
\end{multline}
We note the term $\prod(1-\mu_i\lambda_j )$ is the resultant of $\phi(L)$ and $L^{q}\theta(L^{-1})$ up to a scale factor.

\section{Gradient and Hessian of the likelihood function}
Between the two functions $\mL(\mu,\btheta, \Phi, \Omega)$ and $\mLb(\btheta)$, we mostly deal with the second one in calibration. The Hessian of the first one is related standard errors of the regression coefficients.
We use the notation $\partial_i$ as short hand for $\frac{\partial}{\partial \theta_i}$.
Let $\T(f,T)$ to be the Toeplitz map mentioned above mapping a power series $f$ to its truncated lower triangular Toeplitz matrix. We have
\[ \Theta_T = \T(\btheta,T)\]
We have 
\[ \Theta_T^{-1} = \T(1/\btheta,T)\]
\[ \partial_i \Theta_T^{-1} = -\T(L^i /\btheta^{2},T)\]

\[\frac{\partial\Theta_{*,T-q}}{\partial \theta_i} = \begin{pmatrix} 0_{q-i+1,i-1} & I_{q-i+1} \\ 0_{T-q+i-1,i-1} & 0_{T-q+i-1,q-i+1}
\end{pmatrix}
\]
Here $0_{ab}$ denotes the zero matrix block of size $a\times b$, so the right hand side of the last equation has an identity matrix of size $q-i+1$ on the right top corner and zero everywhere else.

Put $\Theta^{-1}_T$ in block matrix form of  sizes $T\times (q-i +1)$ and $T\times (T-q+i-1)$: 
\[\Theta^{-1}_T =  [(\Theta^{-1}_T)_{T, q-i+1} (\Theta^{-1}_T)_{T, T-q+i-1}] \]
Then
\[  \Theta^{-1}_T \frac{\partial\Theta_{*,T-q}}{\partial \theta_i} = (0_{T,i-1}, \Theta^{-1}_{T,q-i+1})
\]
\[\partial_i \lambda = \partial_i [\Theta_T^{-1} \Theta_{*,T-q}] = (\partial_i \Theta^{-1}) \Theta_{*,T-q} + (0_{T,i-1}, \Theta^{-1}_{T,q-i+1})
\]
\[ \partial_i\bar{K} = (\partial_i \lambda)^{\prime}\lambda + \lambda^{\prime}\partial_i\lambda \]
\[ \partial_i\bar{K}^{-1} = -\bar{K}^{-1}(\partial_i\bar{K}) \bar{K}^{-1}\]

\[\partial_i \bX_{\theta} = -\T(L^i\btheta^{-2} ,T) \bX \]
\[\partial_i \bX_{\theta,\lag} = -\T(L^i\btheta^{-1} ,T)\bX_{\theta,\lag} \]

For any two matrices $A$, $B$ depending ont $\theta$
\begin{multline}
\partial_i AKB = (\partial_i A)KB + AK (\partial_i B) -A(\partial_i \lambda)\bar{K}\lambda^{\prime}B - A\lambda\bar{K}(\partial_i \lambda)^{\prime}B - \\ 
A\lambda (\partial_i \bar{K})\lambda^{\prime}B
\label{eq:Kprod}
\end{multline}
In case $A=B$, the calculation is further simplified as the result is symmetric so we only need to compute half of the terms. Apply (\ref{eq:Kprod}) with $A$, $B$ as $\bX_{\theta}$ or $\bX_{\theta,\lag}$ we can compute the partial derivatives of
\[D(X,\theta) = \bXtheta' K \bXtheta \]
\[C_{\lag}(X,\theta) =\bX_{\theta,\lag}^{\prime}K\bX_{\theta,\lag}\]
\[B_{\lag}(X,\theta) = \bX_{\theta,\lag}^{\prime}K \bXtheta\]

$\bOmega_{opt}$ could be written as $D(X,\theta)-B_{\lag}C_{\lag}^{-1}B_{\lag}$ with $A,B,C,D$ are computed from $X_{\theta}$ and $X_{\theta,\lag}$ so we can calculate its derivatives with the help of
\[ \partial_i (B^{\prime}C^{-1}B) = (\partial_i B')C^{-1}B +  BC^{-1}(\partial_i B)- B^{\prime}C^{-1}(\partial_i C) C^{-1} B.
\]
Putting everything together we have the partial derivative $\partial_i \bOmega_{opt}$. Furthermore
\[\partial_i \log(\det(\bOmega_{opt})) =\Tr(\bOmega^{-1} \partial_i \bOmega_{opt})\]
\[\partial_i \log(\det(\lambda^{\prime}\lambda+ I_q)) = \Tr(\bar{K}^{-1}\partial_i \bar{K})\]
\[\partial_i \mLb(\btheta) = -\frac{T}{2} \partial_i \log(\det(\bOmega_{opt})) - \frac{k}{2}\partial_i \log(\det(\lambda^{\prime}\lambda+ I_q)) \]
So we have the gradient of $\mLb$. 

Applying the chain rule and various matrix derivative rules we can also compute the Hessian of $\mLb$. Here, we will need 
\[
\partial_i\partial_j \Theta_T^{-1} = 2\T(L^{i+j}/\theta^3,T)
\]
and derivative of matrix product rules. The calculation is tedious but doable. However we will not pursue its calculation here. 

From the general theory of Fisher matrix for maximum likelihood estimates, the Hessian of $\mL$ is related to standard error estimates of the $\btheta$ and $\bPhi$. We note the work of Klein and Melard \parencite{KleinMelard} for general (matrix $\btheta$) VARMAX case.

From the expression of $\mL$, the Hessian block $\bH_{\btheta\btheta}(\mL)$ is rather complex and could be done through FFT convolution involving convoluting $1/\btheta, 1/\btheta^2, 1/\btheta^3$ with $X$. We could compute it numerically. We note however the blocks $\bH_{\btheta \bPhi}(\mL)$ and $\bH_{\bPhi \bPhi}(\mL)$ are rather simple, the first one could be computed from (\ref{eq:Kprod}) then replacing one of the two $\bZ$ terms with $\bX$. The second is a direct generalization of the VAR case:
\[\bH_{\bPhi \bPhi}(\mL) = \Tr(\bX \Omega_T^{-1\prime}K\Omega_T^{-1}\bX\Omega^{-1})
\]

Another approach to gradient would be to consider $\mL$ and $\mLb$ as functions of $\Sigma_T$, hence as functions $\gamma_i$, then express $\gamma_i$ as functions of $\btheta$, as done in \parencite{AndersonTakemura}. Their calculations show the interesting fact that the Jacobian $\frac{\partial \gamma_i}{\partial \theta_j}$ looks closely related to the Szeg\"o limit of the determinant of $\Kbar$:
\[ \det (\frac{\partial \gamma_i}{\partial \theta_j}) = \theta_0^{q+1}\prod (1-\lambda_i^{-1}) 
\prod (1+\lambda_i^{-1})
\prod (1-\lambda_i^{-1}\lambda_j^{-1})
\]

Finally, we could extend this approach for the case where the coefficients of $\btheta$ are functions of a finite number of parameters $p_j$. If we deal with gradients only we only need the matrix $(\frac{\partial \theta_i} {\partial p_j})$ then apply the chain rule. So we can apply our core codes to calibrate even more general models. We will discuss this later in sections on extended models.

\section{Computation and Calibration}
We have mentioned the computation of $\Omega_T^{-l}S$ is just a convolution of $1/\btheta^l$ with $S$. The only long matrix calculation encountered is of the form $\T(1/f,T)A$ with $f$ is one of $\btheta, \btheta^2, \btheta^3$ and $A$ is one of $X$ or $\Theta_{*;T-q}$. The convolution calculation could be done through Fast Fourier Transform.

First is the calculation of $1/f$. If $f$ is a polynomial of low degree, which we most likely will encounter, we can either use a recursive algorithm to calculate $1/f$ or expand $f$ to partial fractions of form $c/(1-d L)$ then apply power series expansion to the later. Another method is to use a fast convergent expansion of $1/f$, see for example \parencite{Harvey}.

To compute the convolution using Fast Fourier transform, we assume coefficients of $1/f$ will be small enough to be ignored after $T_c(\btheta)$ steps. FFT convolutions algorithms divide $T$ in to short segments where FFT could be computed efficiently, and make use of the fact convolution is transformed to component-wise multiplication after FFT. The segments are then patched together using the overlap-save method, for example.

In practice, since we also need to compute $\Theta^{-1}_T L^iX$ for $i=0,\cdots p$ it is more convenient to compute $\Theta^{-1}_{T+p} \bhX$ by FFT convolution. As before 
\[
\bhX = \begin{pmatrix}
X_1\\ \vdots \\ X_p \\ \vdots \\ X_{T+p}
\end{pmatrix}
\]
Write $C=\Omega^{-1}_{T+p}$ and $\bhX$ in blocks of size $p-i$, $T$ and $i$:
\[
\Theta_{T+p}^{-1} \bhX = \begin{pmatrix}
C_{[1:(p-i),1:(p-i)]} & 0 & 0 \\
C_{[(p-i+1):(T+p-i),1:(p-i) ]} & \Theta_{T}^{-1} & 0  \\
C_{[(T+p-i+1):,1:(p-i)]} & C_{*} & C_{**} 
\end{pmatrix}
\begin{pmatrix}
\bhX_{[1:(p-i)]} \\
L^i \bX \\
\bhX_{[(T+p-i+1):]}
\end{pmatrix}
\]
We see the block of rows $p-i+1$ to $T+p-i$ of $\Theta_{T+p}^{-1} \bhX$ is 
\[ C_{[(p-i+1):(T+p-i),1:(p-i)]} \bhX_{[1:(p-i)]} + \Theta_{T}^{-1} L^i \bhX \]
From here $\Theta_{T}^{-1} L^i \bX$ is backed out by subtracting these rows by the first term. The submatrix of $\Theta_{T+p}$ could be expressed in term of segments of the power series expansion of $\btheta(L)^{-1}$. This adjustment computation is at cost of $O(T\times i)$ each for a total cost of $O(T p(p+1)/2)$ total. For large $T$ the contribution of the adjustment block decays relatively fast.

We note the inversion of $\btheta$ is $O(T\log(T))$ and the convolution is $O(kT\log(T))$ if we use FFT. Linear regression is $O((kp+1)^2T)$ if $T$ if $T$ is much larger than $k$. Overall, the computation of the likelihood function is of order $O(c_1(kp+1)^2T +c_2 kT\log(T))$ for constants $c_1, c_2$. This is already an improvement over the $O(T^2)$ estimate for standard Kalman filter calculation. See, however \parencite{FastKalman} for an approximation of time $O(T+\log(T))$.

The strength of the method is in calibration. We only need to optimize the function $\mLb(\btheta)$ in the $\theta$ parameter. This function is much simpler to compute and optimize than the traditional  Kalman filter approach. First of all, the function is symmetric, we can save half of the calculation by applying transposes. The matrices encountered here are triangular Toeplitz matrices. Secondly, $K$ is the only large square matrix encountered. But we never need to calculate $K$ directly. Recall 

\[\bar{K} = I_q +\lambda^\prime\lambda \]
$\bar{K}$ is of size $q$, and also symmetric. Therefore we can do a Cholesky decomposition
\[ \bar{K} = C_K C_K^{\prime}
\]
Here $C_K$ is a lower triangular matrix of size $q\times q$.
so all the subsequent calculation are all based on triangular matrices. Since
\[K = I_T - \lambda \bar{K}^{-1} \lambda^{\prime} \]
to compute $N^{\prime}KM$, with $N$ and $M$ has a small number of columns, and of $T$ rows, we actually need to compute $N'M$, $N'\lambda C_K^{-1\prime}$ and $C_K^{-1}\lambda^{\prime}M$. All the matrix multiplications and inversions here involve matrices of size $m\times T$ with $m \ll T$.

The next question is how to determine $p$ and $q$. Here we come back to the earlier discussion on expressing $N(L)^{-1}D(L)$ to the scalar denominator form, with $\btheta(L) = \det(N(L))$ and $\Phi(L)= N_A(L) D(L)$. IF $N$ and $D$ comes from a minimal realization, we see both $\deg(det(N(L))$ and $\deg(N_A(L))D(L)$ are smaller than the McMillan degree $m$. So we can choose to maximize likelihood with both $p$ and $q$ equals to the McMillan degree, then apply tests to determine which higher oder terms could be eliminated, with help of information criteria. This will need further research since this suggestion may be far from optimal if the actual degree of $q$ is much smaller than $m$. We expect cannonical correlation analysis to play a role here.

Let us now discuss the actual maximization of the likelihood function.  We recall again that the likelihood formula is valid for $\btheta$ with roots anywhere in the complex plane. However, if we try to evaluate it with $\btheta$ with at least one root inside the unit disc, both $\Theta_T$ and $\bOmega$ will assume a very large value, even though the likelihood function remains finite. The root invertion maps, discussed in section (\ref{appxIR}) give us in every case a pair $(\btheta_{IR},\bOmega_{IR})$ with the same likelihood function value, where $\btheta_{IR}$ has roots outside the unit disc. The case of roots on the unit cirle is special. We will discuss it briefly in section \ref{appxIR}. 

We can always choose the initial model to be invertible, and we must if we want the system to be identifiable. The region of $\btheta$'s where $\btheta(L)$ has roots outside of the unit disc is a convex connected region, as discussed in the section \ref{appxIR}. The likelihood function is not in general convex. Indeed, a careful analysis of scalar MA($1$) case in \parencite{Davis1} shows it could have several local maxima with different asymptotics. In the next section we discuss some properties of this region, and how to choose the initial points for the calibration. For the actual calibration the reader can choose his favorite gradient-based optimizer, L-BFGS-B is the author's method of choice for this case. We force the optimization to stay within the invertibility region by assigning a large value to the cost function when it is outside. Presumably there may be a better algorithm taking to account the shape of the region as well as the way the gradient transforms under the root inversion map.

\section{Root invertion maps}
\label{appxIR}
Hansen and Sargent \parencite{HansenSargent} (also \parencite{Hamilton} for a detailed exposition) proposed a scheme to transform any MA systems to one with roots within the unit disc. Under that scheme the autocovariance-generating function of the new model is the same as the original one. We show here that these transformations works in the case of VARMA with scalar $\btheta$ and check that they also preserve the conditional log-likelihood. This is a direct generalization of the similar result in the scalar moving average case.

Assuming the equation $\btheta(L)=0$ having roots $\lambda_1^{-1},\cdots, \lambda_q^{-1}$: 
\[ \btheta(L) = \prod_{l=1}^{q}(1-\lambda_lL)\]
Recall $Z_t = \sum_{i=0}^q \theta_i \epsilon_{t-i}$, so $Z$ is a VMA process. Let $\gamma_i$ be given in (\ref{eq:gamma}). Consider the autovariance-generating function of $Z$:
\[ g(z;\btheta;\bOmega) = (\sum_{i=-\infty}^{\infty} \gamma_i z^i ) \bOmega =\prod_{l=1}^{q}(1-\lambda_lz)( 1-\lambda_lz^{-1}) \bOmega\]

Let us recall how the root invertion maps are constructed. Choose a subset of indices $i_1,\cdots, i_r$ correspoding to $\lambda_{i_1}\cdots \lambda_{i_r}$ and consider the polynomial
\begin{equation} \btheta_{IR}(L) =  \btheta_{IR_{i_1,\cdots,i_r}}(L) = \prod_{i\not\in \{i_1\cdots i_r\}} (1-\lambda_iL)\prod_{i\in \{i_1\cdots i_r\}}(1-\lambda_i^{-1}L)
\end{equation}
and the covariance matrix
\begin{equation}
\bOmega_{IR} = (\lambda_{i_1}\cdots \lambda_{i_r})^{2}\bOmega
\end{equation}
The following theorem summarizes some important the properties of this map:
\begin{theorem}
Under the root invertion map corresponding to $i_1\cdots i_r$, the autocovariance generating function of $\btheta_{IR}$ is invariant:
\begin{equation} g(z;\btheta_{IR};\bOmega_{IR}) = g(z;\btheta; \bOmega) 
\label{eq:irautocor}
\end{equation}
Let $\Theta_{IR;T}$ and $\bar{K}(\btheta_{IR})$,  $K(\btheta_{IR},T)$ be the Toeplitz matrix and $K$-matrix corresponding to $\btheta_{IR}$ we have:
\begin{equation} \Sigma_{IR;T} := \Theta_{IR}K_{IR}^{-1} \Theta_{IR}^{\prime} = (\lambda_{i_1}\cdots \lambda_{i_r})^{-2}\Sigma_T
\label{eq:irsigma}
\end{equation}
\begin{equation}
\det(\bar{K}(\btheta_{IR}))  = (\lambda_{i_1}\cdots \lambda_{i_r})^{-2T}\det(\bar{K}(\btheta)) 
\label{eq:irkbar}
\end{equation}

\begin{equation}
\begin{pmatrix} \mu \\ \Phi
\end{pmatrix}_{opt} (\btheta_{IR}) =
\begin{pmatrix} \mu \\ \Phi
\end{pmatrix}_{opt} (\btheta)
\label{eq:irphi}
\end{equation}
\begin{equation}
\bOmega_{opt}(\btheta_{IR}) = (\lambda_{i_1}\cdots \lambda_{i_r})^{2}\bOmega_{opt}(\btheta)
\end{equation}

The conditional likelihood functions are also invariant under this transformation.
\begin{equation}
\mL (\btheta_{IR}, \mu, \bPhi, \bOmega_{IR}) =\mL (\btheta, \mu, \bPhi, \bOmega)
\label{eq:irllk}
\end{equation}
\begin{equation}
\mLb (\btheta_{IR}) =\mLb (\btheta)
\label{eq:irllkb}
\end{equation}
\label{eq:lemriv}
Let $J[IR]$ be the gradient of $IR$. The gradient of $\mLb$ transforms under:
\[ \nabla \mathscr{L}(\btheta)  = \nabla \mathscr{L}(\btheta_{IR}) J[IR]
\]
\end{theorem}

The proof of (\ref{eq:irautocor}), (\ref{eq:irsigma}) are the same as the scalar case in \parencite{Hamilton}. (\ref{eq:irphi}) is straight forward from its expression. (\ref{eq:irkbar}) follows from $\det(\bar{K})=\det(\Sigma_T)$. The last two equations come from direct substitutions.

Note $\lambda_1,\cdots,\lambda_q$ are roots of $L^q\btheta(L^{-1})$, we may sometime call them roots unless there is confusion. The invertibility condition is $|\lambda_i| \leq 1$.

We have the Vieta map $\Vt$ from roots $(\lambda_1,\cdots \lambda_q)$ to coefficients $(\theta_1,\cdots \theta_q)$. This map is well defined and algebraic, given by symmetric polynomial equations:
\[\theta_l = \sum_{i_1,\cdots,i_l} (-1)^l\prod_{i \in {i_1,\cdots,i_l}}\lambda_{i}\]

The map $\Vt$ is not one-to-one, at least any permutation of $\lambda_1,\cdots,\lambda_q$ give us the same coefficients. Now the root inversion maps described above are defined on the root space, but only defined on the coefficient space after we have chosen a partial inverse of $\Vt$, which is a particular ordering of the roots. With this in mind we will determine the effect of root inversion on the gradient of the likelihood function. We will use the chain rule and the implicit function theorem and for that we will need the Jacobian of $\Vt$
\[J_{\Vt} = (\frac{\partial \theta_i}{\partial \lambda_j})
\]
\[ \frac{\partial \theta_i}{\partial \lambda_j} = (-1)^{i-1}\sum_{(\boldsymbol{l}): |(\boldsymbol{l})| = i-1} \prod_{j\not\in (\boldsymbol{l})} \lambda_{l\in (\boldsymbol{l})} \]
Which denote sum over products of $i-1$ elements that does not contain $j$. So the $j$th column is just the coefficients of the expansion of $-\btheta(L)/(1-\lambda_jL)$. A trick we use repeatedly is to evaluate complex expressions at roots of unity, then apply IDFT to compute the coefficients. In the code we use this trick to evaluate the later expression. We note that the Jacobian could be complex if some of the roots are complex. Also $J_{\Vt}$ is not invertible at roots with multiplicity, and the inverse function is not well defined there.

$IR_{i_1,\cdots,i_r}$ is considered as a map from the root space to itself, sending $(\lambda_{i_1}\cdots \lambda_{i_r})$ to their inverse. For it to act on the coefficient space, we need to solve the equation $\btheta(L) = 0$, take the inverse of the $\lambda_{i_1} \cdots \lambda_{i_r}$ then reconstruct the coefficients. In effect it is $\Vt \circ IR \circ \Vt^{-1}$. The chain rule and the implicit function theorem gives

\[ J[IR_{i_1,\cdots,i_r} (\btheta)] = J_{\Vt}|_{\Vt^{-1}(\btheta_{IR})} \text{diag}(1,\cdots -\lambda_{i_1}^{-2},\cdots, -\lambda_{i_r}^{-2}\cdots,1) J_{\Vt}^{-1}|_{\btheta}
\]

We note that we need to pick $S=\{\lambda_{i_1}\cdots \lambda_{i_r}\}$ so that if $\lambda_{i}$ is in $S$ then $\bar{\lambda}_i$ is also in $S$. In that case $J_{IR}$ is real, as it is the Jacobian of a real map, even if $J_{\Vt}$ could be complex.

Finally by the chain rule and invariance of $\nabla \mathscr{L}(\btheta)$ under the action of $IR$ gives us the equation for the gradient.

Note if $f(\btheta_{IR})$ transforms as
\[ f(\btheta_{IR}) = h(\theta)f(\btheta) \]
where $h$ is a scalar function and $f$ is a vector function 
then we have 
\[  \nabla f(\btheta)  =\frac{1}{h(\btheta)} \nabla f(\btheta_{IR}) J[IR] -\frac{1}{h(\btheta)^2} (\nabla h)_{|\btheta}  f(\btheta_{IR})
\]
If $h$ is given in term of $\lambda$, for example $g(\lambda) = \lambda_{i}^2$ then $\nabla h$ could be computed using the Jacobian of the Vieta map
\[ \nabla_{\btheta} h = \nabla g_{\lambda} J_{\Vt}^{-1}\]
and from here we can also compute $\nabla f(\btheta)$. In practice we only need to compute the gradient of $\mLb$, but it is useful for sanity check to compute gradient of the intermediate terms.

With these relations, we can compute both the value and the gradient of the likelihood function at a non-invertible $\btheta$ by transforming it to an invertible point where the calculation is numerically stable. Hence we can apply gradient optimization method without any restriction on the domain of $\btheta$. Note that we will work with $\btheta$ that has no multiple roots here where $J[IR]$ is defined.

The root inversion maps have some interesting property near it fixed points as seen in the next lemma.
\begin{lemma}
If $\btheta$ is fixed under a root inversion map $IR=IR_{i_1\cdots i_r}$ then 
\begin{equation}
J[IR]^2(\btheta) = I_q
\end{equation}
In that case, $J[IR]$ has eigenvalues of $-1$ or $1$ only. 
More over we have
\begin{equation} \nabla \mLb(\btheta)J[IR]  = \nabla \mLb(\btheta)
\end{equation}
\end{lemma}
The first statement is a consequence of the fact that $IR \circ IR=id$ around $\btheta$. The second is clear from invariance of the action of $IR$ on the likelihood function and the fact that $\btheta$ is a fixed point.

We see that this puts constraints on $\nabla \mLb$. If we split the tangent space of $\btheta$ at a fixed point of $IR$ to eigenspaces of $J[IR]$, corresponding to eigenvalues $\pm 1$ there is no constraint on the eigenspace corresponding to $1$, while if $c$ is an eigenvector corresponding to $-1$ then $\nabla\mLb . c = 0$. For MA($1$) this is already well-known, as $J[IR]=-1$ in that case. It is surprising to us that we can do quite a bit better by examining the eigenvalues of the Jacobian in details. In the paper \parencite{JorJac} we prove that the dimension of the eigenspace corresponding to $-1$ is 
\begin{equation}
\text{ mult}_{-1} J[IR](\btheta) = \left\{ \begin{array}{l l} \floor*{r/2} +1 & \text{if } \psi_r = -1 \text{ or } $r$ \text{ is odd } \\
\floor*{r/2} & \text{otherwise}
\end{array}
\right.
\end{equation}
Here, $\psi_r = (-1)^r\lambda_{i_1}\cdots\lambda_{i_r}$ and $\floor*{x}$ denote the integer part of $x$. From here, we see the only cases were $J[IR]-I_q$ is invertible are $q=1$, $\btheta = 1\pm L$ and $q=2$, $\btheta = 1 -  L^2$. Those cases are the cases where the constraint are strongest, the corresponding models are critical points of the likelihood function regardless of the sample data set. This is the pile-up effect. In optimization when we observe a critical point close to these values, additional analysis would be required. On the other hand \parencite{Davis1} has studied the both local and global maximum of the likelihood function in detail for MA($1$) case. Testing for MA unit root has attracted the attention of several authors, see \parencite{AndersonTakemura, Tanaka, Davis1, Davis2011} for the pure MA case. In the later works for MA($1$) case, a change of parameter of form $\theta_1 = 1-\beta/T$ expresses the likelihood function as a function of $\beta$, which could have more than one local maximum point. Tests for MA unit roots could be derived from that study. The analysis make use of a join distribution of the Gradient and Hessian with respect to the changed variable $\beta$. 

Our analysis suggests that when $q$ is larger, the MA unit root constraints are not as strong as when $q$ is small. In the generic cases (corresponding to the hyperplane boundary) where we have one or two conjugated unit roots, we have only one unit root we have at most one constraint on gradient of the likelihood function. At the more complex boundary point the number of constraints is around half the number of unit roots. The constraints could be given very explicitly in term of the unit roots, as we will see in the paper \parencite{JorJac}.

We hope the results here provide some help in analyzing unit roots in general case. This topic requires further studies.

\section{Invertibility region and initial values}
Many results in this section is well-known in the system and control literature. We recall them here for the reader's convenience.
\begin{theorem}
The set $\theta_1,\cdots \theta_q$ such that the equation:
\[ 1+\theta_1 L +\cdots \theta_q L^q =0\]
have roots outside of the unit disc or equivalently the equation:
\[ z^q+\theta_1 z^{q-1} +\cdots \theta_q =0\]
have roots inside the unit disc is a convex, connected set bounded by real, algebraic hyperplanes given by the Schur-Cohn polynomial inequalities.
\end{theorem}

We refer the readers to the literature \parencite{KreinNaimark, Schur, Cohn, Jury, Bistritz} for this classical result and improvements. We do not need the Schur-Cohn boundary explicitly, as it is not too expensive to calculate the roots directly and compare the modulus with one. For readers who are not interested in the details, it is sufficient to know that there exist inequalities formed by algebraic polynomials called the Schur-Cohn polynomials such that the statbility restriction on roots are satisfied if and only if these inequalities are satisfied. The Schur-Cohn polyonomials could be computed recursively via efficient algorithms in the above references. We will show only a few examples for $q\leq 3$ to illustrate the idea. We note for one variable the condition is simply $-1\leq \theta_1 \leq 1$, for two variables the condition is 
\[\theta_2 < 1 \]
\[ -\theta_1+\theta_2 +1 \geq 0\]
\[ \theta_1+\theta_2 +1 \geq 0\]
which form a triangle with (inverse) base $\theta_2=1$ and top at $(0,-1)$. For three variables the Schur-Cohn conditions are

\[ 1+\theta_1 +\theta_2 +\theta_3 > 0\]
\[ 3+\theta_1 -\theta_2  -3 \theta_3 > 0\]
\[1-\theta_1+\theta_2-\theta_3 > 0 \]
\[ 1-\theta_2 -\theta_3^2 +\theta_1\theta_3 > 0\]

The first three conditions give a tetrahedral with vertices $(-1,3,3)$, $(1,-1,-1), (1,3,3),(-1,-1,1)$. The last equation restricts it further to a convex region of the tetrahedral. We note the second condition does not appear in the limit determinant of $\Sigma_T^{-1}$ discussed above. Our simulation shows it is in fact not needed, it seems to be a consequence of the remaining three conditions. We note in our previous discussion of the determinant of $\Sigma_T^{-1}$, the factor $\prod (1-\lambda_i \lambda_j)$ is symmetric and could be expressed as a polynomial in $\theta_i$'s. This function vanishes whenever we have conjugated unit roots so should be closely related to the invertibility boundary. Up to $q=3$ this seems to be the only non linear condition. For $q=4$ that factor is of degree $12$ in $\lambda_i$, while the Schur-Cohn polynomials are of degree at most 6, so the picture is more complex here. It would be nice to understand more clearly the relationship between the Szeg\"o determinant limits and the Schur-Cohn boundary.

While the invertibility region is convex, in general the likelihood function is not, therefore we need to deal with local minima. Here, the cost function is minus the log-likelihood. We have briefly discuss the situation with root on the unit circle in the previous section so in this section we will focus on optimization technique inside the region. While more theoretical work will be needed to understand the distribution of local minima, our first attempt is to use local optimizers with initial points starting in different sub regions, with the hope that when the mesh of sub regions is fine enough we will catch the global optimum point.

Of course there are many ways to choose the starting points, we describe here the method that we use in our code. Recall that a real polynomial of odd degree always have at least one real root, and in general complex roots always appear in conjugated pairs. We look at inverse of roots of $\btheta$, which are roots of $L^q\btheta(L^{-1})$. We are assuming they are inside the unit disc. We will call them roots here when there is no confusion.

Our strategy is for real roots, divide the interval $[-1,1]$ in to regions, and for complex root divide the upper unit disc  in to regions, then consider possible arrangements of the $q$ roots to these regions.

To illustrate, let us divide the interval  $[-1,1]$ to three subintervals: $R_1= (-1,-3^{-1/2}],R_2= [-3^{-1/2},3^{-1/2}], R_3 [3^{-1/2},1)$. We divide the upper half disc to three regions: $C_1$ is the half disc with radius $3^{-1/2}$, $C_2$ is the part of the first quadrant with radius between $3^{-1/2}$ and $1$, and $C_3$ is the part of the second quadrant with radius between $3^{-1/2}$ and $1$. The choice of $3^{-1/2}$ is so that the three complex regions to have the same area, and there is exact overlap between the real and complex regions. We  could modify the choices some other ways.

Set $q = q_r +2q_c$, where $q_r$ is the number of real roots and $2q_c$ is the number of complex roots. Consider the arrangements of the $q_r$ real roots to $q_r = q_{r_1} +q_{r_2} +q_{r_3}$ corresponding to the three interval $R_1,R_2,R_3$ and the $q_c= q_{c_1}+q_{c_2}+q_{c_3}$ complex roots in the upper half plane to the three area $C_1,C_2,C_3$ with $q_{c_1}$ roots in $C_1$,  $q_{c_2}$ roots in $C_2$ and $q_{c_3}$ roots in $C_3$.

We can see the number of choices is $\frac{(q_r+1)(q_r+2)}{2}$ for the real roots, and $\frac{(q_c+1)(q_c+2)}{2}$ for the complex roots. So the number of regions under this partition is
\[
\sum_{q_c \leq \text{floor}(q/2); q_r = q-2q_c } \frac{(q_r+1)(q_r+2)}{2} \frac{(q_c+1)(q_c+2)}{2}
\]

It turns out the sum could be simplified to polynomials of degree five depending on $q$ odd or even: 
\begin{equation}
\text{number of regions =}\begin{array}{cc} 
\frac{(q+2)(q+4)(q+6)(q+8)(2q+5)}{1920} & \text{if $q$ is even}\\
\frac{(q+1)(q+3)(q+5)(q+7)(2q+13)}{1920} & \text{if $q$ is odd}\\
\end{array}
\end{equation}

Start with one region, for example we choose say $q_r =q+0+0$ roots on in $R_1$ and no complex root ($q_c=0$). The roots could be picked randomly or deterministically. For example we will choose them to be just the middle point of $R_1$. Then we construct $\btheta$ from roots by the Vieta formula.

The resulting $\btheta$ will be an initial value for the first local optimization. We repeat this for all regions to choose initial points. While the number of initial points growths polynomially, with our algorithm we can compute the likelihood functions relatively fast for practical data size. In practice we pick the best initial points and optimize them further with a local optimizer.

\section{Additional topics}
\subsection{Seasonality and Integration}
\label{secInt}
First we note the whole process work if we add additional drift terms, or additional regressions. For example to allow a polynomial drift we add vectors of form $i^k$ instead of $1$ in the definition of $\bX_{\lag}$. Seasonality could be accounted for by seasonal dummy variables, just like the VAR case. We will next discuss integrated models. Consider the following model with scalar $\btheta$:
\[ \Phi(L) X = \btheta(L) \epsilon
\]
We note the polynomial division algorithm works for any matrix polynomial and a scalar polynomial. In particular, apply polynomial division of $\Phi$ to $L(L-1)$, note that the remainder matrix is a matrix polynomial of degree at most one we have
\[
\Phi(L) = L(L-1)\Gamma(L)t + \Phi_b(1-L) -\Pi L
\]
($\Phi_b(1-L) -\Pi L$ is the remainder of the division by $L(L-1)$ which is of degree $1$ so will be of form $A+B L$, and we set $\Phi_b = A$, $\Pi = -A-B$).

Let $L=0$ and $L=1$, respectively we get:
\[
\Phi_b = I_k
\]
\[
\Pi = - \Phi_L(1)  = -I_k +\Phi_1+\cdots+\Phi_p
\]
Let $\Delta = 1-L$. The equation becomes:
\[
\Delta \bX(t) =  \Gamma(L)\Delta \bX(t-1) +\Pi X(t-1)+\btheta(L)\epsilon(t )
\]

Apply $\btheta(L)^{-1}$ to both sides we get 
\[
\Delta \bX_{\theta,t} =  \Gamma(L)\Delta L\bX_{\theta,t}  +\Pi L\bX_{\theta, t}+\epsilon_t
\]
where $X_{\theta,t}$ is $\theta(L)^{-1}X(t)$. This is our VECM form. We can apply an argument similar to Johansen for cointegration here. We construct a regression between $\Delta \Theta_T^{-1}\bX_t$ and the lags represnted by $\bX_{\theta,\lag}$, where $\bX_{\theta,\lag}$ consists of terms  $\Delta \Theta_T^{-1}L^1\bX_{t-1},\cdots \Delta \Theta_T^{-1}L^p\bX_{t-p+1} $  and $\Theta_T^{-1}L\bX_{t}$. This is essentially the same construction of the VARMA case, the integration component correspond to the term $\Theta_T^{-1}L\bX$.  Let  $r \leq k$ be the rank of $\Pi$ . If the $r =k$ then we have a stationary process. If $r=0$ we do not have cointegration. If $0< r <k$ then we have a cointegrating system. We can decompose $\Pi = \alpha \beta^T$ with $\alpha, \beta$ are a $k\times r$ matrix of full rank. It remains to apply a rank test to figure out the rank $r$. We expect a result similar to Johansen's test \parencite{Johansen1991} where the inner product defined by $\Sigma_T$ plays a role.

\subsection{Extension to infinite component MA}
Next, a few words about the case when we have an infinite number of MA components. If we aim to study models with a finite number of VAR terms but an infinite MA scalar terms, we expect the result here to carry through, provided we apply the appropriate inner product constructed from the MA scalar terms. So the issue is to study this inner product. The survey paper \parencite{Bingham2012} provided a framework to think about the MA($\infty$) case. Blaschke product used by Hansen and Sargent is closely related to Hardy spaces, so it has been understood for sometime that Toeplitz operators, Wiener-Hopfs, Hardy spaces are what needed to extend the theory to infinite component moving average models. In a future paper we hope to work out the technical details.

Since we deal with an infinite past, a rigorous approach may require more analytic machinery than we intent to cover here, but let us sketch a few ideas. When we have infinite MA terms, the invertible condition is just the condition that $\btheta(L)$ has no root or pole inside or on the unit circle (see the next section for the case of poles on the unit circle - as in case of fractional Gaussian). $\btheta$ is called an {\it outer function} or {\it Szeg\"o function}. An example of such function could be any stable $ARMA$ rational function (we presumably formulate that all entries of $\bPhi(L)$ has a polynomial factor $f$ and use $f/\btheta$ as our MA($\infty$) function). A function of form $(1-b_1 L)^{\alpha_1}\cdots (1-b_m L)^{\alpha_m}$ with $|b| <1$ is also an outer function. We expect to be able to apply our framework to calibrate a finite number of parameters that generate a model with infinite moving average components.

$\Sigma_T$ and $K$ are finite dimension but now $\bar{K}_T$ is of infinite dimension. We will need a definition of Gaussian measure as well as determinant in this context - both of which fortunately have been studied for a long time. The discussion of infinite dimensional MA with analytic outer function would hopefully provide error estimates to our main regression of $\Phi$ when we cut off the expansion of $\btheta$ by a finite number $q_c(\btheta)$ of terms.

Let us shift the index by $1$ and consider the index set of the sample as $\{0,\cdots, T-1\}$ instead of $\{1,\cdots, T\}$. This makes it more convenient when we write convolutions.
Set $\btheta_+(L) = \btheta(L)$ and $\btheta_{-}(L) = \btheta(L^{-1})$, considered as Laurent series. Consider the vector space $V_{\geq 0}$ spaned by basis $\{v_i\}_{i=0}^{\infty}$. For any Laurent series $a(L) = \sum_{-\infty}^{\infty}a_i L^i$ define the infinite Toeplitz matrix $\T_{\infty}(a) = (a_{j-k})_{j,k=0}^{\infty}$. This is the matrix of the action of convolution of $a(L)$ on $V_{\geq 0}$:
\[a.v_k = \sum_{j=0}^{\infty} a_{j-k} v_{j} \]
(we will need a norm for the sum to make sense). In our paper we work with the top $T\times T$ block of this matrix. We note $\T_{\infty}(\btheta_+)$ is the infinite version of $\Theta_T$, $\T_{\infty}(\btheta_-)$ is the infinite version of $\Theta^{-1}_T$, $\T_{\infty}(\btheta_+\btheta_-)$ is the full autocovariance matrix, and its upper left $T\times T$ matrix is our $\Sigma_T$. In their second proof of the Borodin Okounkov's formula \parencite{BasorWidom}, the authors defined a matrix $A$ as
\[A = \T_{\infty} (\btheta_+^{-1})\T_{\infty}(\btheta_+ \btheta_-) \T_{\infty}(\btheta_-)
\]
and showed
\[ A^{-1} = \T_{\infty}(\btheta_-\btheta_+)\T_{\infty}(\btheta_+\btheta_-)\]
we note $K(\btheta, T)$ is just the upper left $T\times T$ block of $A^{-1}$. They noted that $A^{-1}-I$  is of trace class. We have shown earlier in case $\btheta(L)$ is polynomial this trace class part is $\lambda\lambda^{\prime}$. To define $\lambda$ in the MA($\infty$) term case we will need some analysis tools which we will not get in to in this paper but formally we can mimic the definition of the polynomial case and define it as an infinite dimensional matrix. We note $\Theta_*$ now acts on an infinite dimensional Hilbert space corresponding to $\epsilon_{i<0}$ (note we shifted the indices by $1$), $\Theta_T$ is defined as before and $\lambda=\Theta_T^{-1}\Theta_*$ is a linear operator represented by a matrix with columns indexed by negative integers and row indexed by $\{0,\cdots,T-1\}$. The interested reader could work out the AR($1$) case where $\btheta(L) = \sum_{i=0}^{\infty} \phi^i L^i$ and find $\lambda =(\lambda_{ij})_{i=0,j=-\infty}^{i=\infty,j=-1}$ with $\lambda_{0j}=\phi^{-j}$ and $\lambda_{ij}=0$ with $i\neq 0$. From here $(\lambda\lambda^{\prime})_{ij} =\frac{\phi^2}{1-\phi^2}$ if $i=j=0$ and zero otherwise, and get to the exact likelihood function of AR($1$).

We see the determinant of $\Sigma_{\infty}$ could now be expressed in two different ways, $\det(I+\lambda\lambda^{\prime}) =\det(I+\lambda^{\prime}\lambda)$. \parencite{BasorWidom} showed the first determinant is the same Fredholm operator determinant in Borodin-Okounkov's formula. In either AR or MA case we expect one of the determinants to collapse to a finite dimensional determinant, but in general we have two Fredholm operator determinant expressions of $\det{\Sigma_T}$. We note that if the coefficients decay sufficiently after $q_c < T$ terms, we only need $q_c$ MA terms in the second expression. We note although we have an infinite (or $q_c$) number of MA terms, in general they are controlled by a finite (and smaller than $q_c$) number of parameters and the gradient calculation would also apply with appropriate application of the chain rule. We expect our calibration method would still be effective in this last case, however to be practical the models need to be in special forms for us to check the invertibility condition.

\subsection{Fractional VARMA}
We again assume finite dimensional VAR model, with a fractional Gaussian MA component 
\[ (1+\Phi_1 L+\cdots \Phi_p L^p )(1-L)^d X_t = (1+ \theta_1 L +\cdots \theta_qL^q)\epsilon_t\]
Here the scalar function to consider is
\[\btheta_{d}(L) =(1-L)^{-d}\btheta(L) \]
which could be written in MA($\infty)$ form. We conjecture the main theorem is still valid in the form given by the matrix $\Sigma_T$ which is finite dimensional, however careful analysis is needed to define $\lambda$, as seen in the previous section. We note the determimant $\Sigma_T$ tends to infinitive at large $T$. If we apply mechanically the Szeg\"o limit theorem we see beside the inverse polynomial terms, the determinant $\det(\Sigma_T)$ would have an extra term corresponding to $d\log(1-L)$:
\[ \exp(\sum_{k=1}^T  \frac{d^2}{k})\]
which increases as $T^{d^2}$. This is a special case of the Fisher-Hartwig conjecture \parencite{FisherHartwig} which has been proved for some time \parencite{Ehrhardt}. In fact in Toeplitz operator literature, people consider function with several (conjugated) poles on the unit circle, as well as other types of singularities. The analysis near singular/zero points on the unit circle would need careful analysis, and we hope operator theory method to be helpful here.

We note that while there need to be theoretical justifications, invertibility considerations and initial point selection, for the last few sections, the algorithms and coding require little modifications. As these models are dependend on a finite set of parameters $p_i$, we only need functions to supply the coefficients $\theta_i$ and the gradient matrix $\frac{\partial \theta_i}{\partial p_j}$, which will be model dependent.

\section{Conclusion}
We have tested the likelihood function and calibration algorithm presented here in R and C++ codes. The philosophy of replacing the scalar MA components with an inner product defined by the finite Toepliz matrix seems fruitful and we expect may other results related to Vector Auto Regressive models are to have corresponding VARMA analogues. It remains to be seen how the calibration algorithm suggested here applies in practical forecast.

\begin{appendices}
\appendix
\counterwithin{lemma}{section}
\section{A few matrix facts}
\label{appxMat}
\begin{lemma}
Let $X, Y,\beta,K,\Omega$ be matrices with compatible dimension such that the following expression is well formed
\begin{equation}
\Tr((Y^\prime-\beta^{\prime}X^{\prime})K(Y-X\beta)\Omega)
\end{equation}
Assume further, that $K$ and $\Omega$ are invertible symmetric positive definite matrices. Also assume $(X^{\prime} K X)$ is invertible. With $X, Y, K, \Omega$ known, the above expression has its minimum at
\[ \beta_{opt} = (X^{\prime}K X)^{-1} X^{\prime} K Y \]
and thus $\beta$ is independent of $\Omega$.
\label{lem:cauchy}
\end{lemma}
Proof. Set $\beta = \beta_{opt}+b$ and expand the expression.
\begin{multline}
\Tr((Y^\prime-\beta^{\prime}X^{\prime})K(Y-X\beta)\Omega) = \Tr((Y^\prime-\beta^{\prime}_{opt}X^{\prime})K(Y-X\beta_{opt})\Omega) - \\
\Tr(b^{\prime}X^{\prime}K(Y-X\beta_{opt})\Omega) - \Tr((Y^\prime-\beta^{\prime}_{opt}X^{\prime})KXb\Omega) + \\
\Tr((b^{\prime}X^{\prime}KXb)\Omega)
\end{multline}
Now note
\[
(X^{\prime}K X)\beta_{opt} =  X^{\prime} K Y 
\]
\[
\beta_{opt}^{\prime}(X^{\prime}K X) =  Y^{\prime} K X
\]
We see both middle terms are zero, while the first and last terms are positive because Kronecker product of positive definite matrix $K$ and $\Omega$ is also positive definite and applying lemma \ref{lem:traceKro}. So the minimum is attained at $b=0$.

This proves the optimality of $\begin{pmatrix} \mu \\ \Phi\end{pmatrix}_{opt}$.

The following is already well-known:
\begin{lemma}
Assuming $K$ is positive definite. Then
\begin{equation}
P_K = K - KM(M^{\prime}KM)^{-1}M^{\prime}K
\end{equation}
is positive semi-definite for all $M$ such that $M^{\prime}KM$ is invertible.
\end{lemma}
Proof: Consider a decomposition $K = L^{\prime}L$ and set $LM=M_1$. We see
\[
L^{\prime -1}P_K L^{-1} = I_T - LM(M^{\prime}L^{\prime}LM)^{-1}M^{\prime}L^{\prime} =I_T- M_1(M_1^{\prime}M_1)^{-1}M_1^{\prime}
\]
is a projection, and hence has eigenvalues $0$ and $1$. So $P_K$ is positive semi-definite.

So we have in particular $\bOmega_{opt}$ is positive semi-definite, regardless of the sample data $\bX$. In practice there may exist data $\bX$ such that $\Omega$ has a zero-eigenvalue. Some regularization need to fix $K$ for that case.

\section{Simulation results}
Using our R script we have tested and confirmed the relationship between $\Sigma_T$ and $K$ and $\Kbar$. We also have confirmed the Szeg\"o limit of the determinant. We also have confirmed the invariant of the likelihood function under the root inversion algorithm, this was done against small sample as large sample data would lead to implosion in intermediate steps.

Using simulated data then maximizing the likelihood function we are also able to recover original models in our test cases. This includes 
\begin{itemize}
\item simple $p=0, q=2$ models (with $2\times 2$ scalar $\btheta$.

\item ARMA $p=2,q=2$ model

\item VARMA models with $k=2, (p=1,q=1)$ matrix polynomials:
\[  \Phi_1 = - \begin{pmatrix}
0.02284288  & 0.4027705 \\
1.06073525  & -0.2589487 \\
\end{pmatrix}
\]
\[
\Theta_1 = -
\begin{pmatrix}
-0.4100472  & 0.3227580 \\
2.1013041  & 0.2378265
\end{pmatrix}
\]
and $X_t = (I - \Phi_1 L) X_t + (I +\Theta L)\epsilon_t$. We write the respective matrix polynomials $ \Theta^M(L)$ and $\Phi^M(L)$. (The long decimals in the matrices were due to the fact we ran a simulation to search for stable matrices.) This is equivalent to a $p=2,q=2$ model with scalar theta $=\det(\Theta^M(L)$, and the AR term $\text{ adj}(\Theta^M(L)\Phi^M(L)$. We are able to recover both the scalar denominator and the degree 2 matrix polynomial numerator.

\item VARMA model with $k=2$ given by ($p=2,q=2$) matrix polynomials. This is equivalent to $(p=4,q=4)$ scalar MA model. Again we recovered the equivalent scalar-denominator model.

\item VARMA model with $k=4$ with $p=5,q=3$ where the numerator is a matrix polynomial and the denominator is a scalar polynomial. While in the previous two cases, we were able to find the optimal parameters by an optimization with initial vector at $0$, for the last case we had to apply the partition of the invertibility region mentioned above.

\end{itemize}
\end{appendices}

ACKNOWLEDGEMENT. We would like to thank Thong Nguyen for very helpful suggestions and help with literature. He filled our gap in knowledge in time series and statistics through insightful conversation and providing reading material. He pointed out (\ref{eq:ylk}) is essentially the Yule-Walker equations, with a different twist.
We thank Utkarsh Samant for encouragement and providing infrastructure where much of test was carried out. Any error remains our responsibility alone.

\printbibliography[title={Bibilography}]

\end{document}